\documentclass[12pt]{article}
\usepackage[letterpaper, left=0.75in, right=0.75in, top=0.75in, bottom=0.75in]{geometry}
\usepackage{graphicx} 
\usepackage{caption}
\usepackage{physics}
\usepackage{amsmath}
\usepackage{subcaption}
\usepackage{booktabs}
\usepackage[sort&compress, numbers]{natbib}
\usepackage{tkz-graph}
\usetikzlibrary{calc} 
\GraphInit[vstyle=normal]
\usepackage{url}
\usepackage{authblk}
\title{Efficient and Noise-Resilient Molecular Quantum Simulation with the Generalized Superfast Encoding}
\author[1]{James Brown}
\author[1]{Tarini S Hardikar}
\author[1]{Kenny Heitritter}
\author[1]{Kanav Setia}
\affil[1]{qBraid Co., Chicago, IL, United States}
\date{October 2025}

\begin{document}

\maketitle
\begin{abstract}
    Simulating molecular systems on quantum computers requires efficient mappings from Fermionic operators to qubit operators. Traditional mappings such as Jordan–Wigner or Bravyi–Kitaev often produce high-weight Pauli terms, increasing circuit depth and measurement complexity. Although several local qubit mappings have been proposed to address this challenge, most are specialized for structured models like the Hubbard Hamiltonian and perform poorly for realistic chemical systems with dense two-body interactions. In this work, we utilize a suite of techniques to construct compact and noise-resilient Fermion-to-qubit mappings suitable for general molecular Hamiltonians. Building on the Generalized Superfast Encoding (GSE) and other similar works, we demonstrate that it outperforms prior encodings in both accuracy and hardware efficiency for molecular simulations. Our improvements include path optimization within the Hamiltonian’s interaction graph to minimize operator weight, introduction of multi-edge graph structures for enhanced error detection without added circuit depth, and a stabilizer measurement framework that directly maps logical terms and stabilizers to the Z-basis using Clifford simulation. Applying these methods to simulations of $(H_{2})_{2}$ and $(H_{2})_{3}$ systems yields significantly improved absolute and correlation energy estimates under realistic hardware noise, with further accuracy gains achieved by increasing code distance. We also propose a [[2N, N, 2]] variant of GSE compatible with square-lattice and (quasi-) linear hardware topologies, demonstrating a twofold reduction in RMSE for orbital rotations on IBM Kingston hardware. These results establish GSE as a very attractive mapping for molecular quantum simulations.  
\end{abstract}

\section{Introduction}
When simulating chemistry on quantum computers in second quantization, it is necessary to map the Fermion operator to qubit operators. This can create high-weight Pauli operators, which results in large circuit depth and can make measuring observables (such as with classical shadows\cite{Huang2020}) more challenging. To overcome this, various local qubit mappings \cite{Derby_2021, Verstraete_2005, Ball_2005, Miller_2023, JKMN, Vlasov_2022, Setia_2019, landahl2023logicalfermionsfaulttolerantquantum, chiew2025optimalfermionqubitmappingsquadratic} have been proposed, which add additional qubits that generally reduce the average weight of the terms in the Hamiltonian. However, these mappings are generally tailored to specific simpler Hamiltonians, such as the Hubbard model with only diagonal 2-body terms and nearest-neighbor interactions. For molecular systems, the Hamiltonian generally includes many two-body terms, which complicate these tailor-made procedures\cite{Chien_2019}. Therefore, it is necessary, when studying chemical systems, to develop procedures that can be utilized for complicated fermionic Hamiltonians.

In this paper, we outline many tools that can be used to generate compact and noise-resilient mappings to simulate general chemical systems more efficiently and accurately on quantum computers. Many of the tools outlined here are not new, but when combined reorient GSE as a very competitive mapping for general fermionic systems. This is in opposition to the original work\cite{Chien_2019} that examined superfast encodings\cite{Setia_2018} for molecules. This did not show any benefit on any important metric compared to the Jordan-Wigner mapping. We show that when utilizing the tools outlined here, the Generalized Superfast Encoding\cite{Setia_2019} (GSE) is a very competitive mapping to use for molecular systems due to the lower circuit depth and improved error mitigation properties that result. 

To exemplify this result, we apply the techniques developed here to the orbital rotation of four electrons in eight orbitals on IBM Kingston using a newly developed [[2N, N, 2]] error detecting code derived from the GSE. The sampled occupancies have approximately half the RMSE error compared to the Jordan-Wigner mapping on linearly connected hardware.

\section{The molecular Hamiltonian}
For chemical systems, it is always possible to use chemist ordering such that there are no `hops' (i.e. $a_i^{\dagger}a_j$) between $\alpha$ and $\beta$ modes. This is done by ordering the modes so that all $\alpha$ modes are followed by all $\beta$ modes (such that the Hamiltonian is in ``chemist'' ordering) which results in the Hamiltonian
\begin{equation}
    H = \sum_{\sigma}\sum_{i,j<M}h_{ij} a^{\dagger}_{\sigma,i}a_{\sigma,j}+\sum_{\sigma,\sigma^{\prime}}\sum_{i,j<M,k,l<M}V_{ijkl}a^{\dagger}_{\sigma,i}a_{\sigma,j}a^{\dagger}_{\sigma^{\prime},k}a_{\sigma^{\prime},l}.
\end{equation}
where $M$ is the number of molecular orbitals and $\sigma$ is the spin. The one-body ($a^{\dagger}_{\sigma,i}a_{\sigma,j}$) term include contributions from the kinetic energy and nuclei-electron terms. The two-body ($a^{\dagger}_{\sigma,i}a_{\sigma,j}a^{\dagger}_{\sigma^{\prime},k}a_{\sigma^{\prime},l}$) term represents the electron-electron interactions. 

This ``chemist'' ordered Hamiltonian only has direct interactions (i.e. movements of particles $a_i^{\dagger}a_j$) within a spin-sector. Therefore, the interaction graph is generally (for molecules) fully connected within a spin-sector but with no connections between spin-sectors. These interactions are represented as edge ($A_{ij}$) and the occupations of those modes by vertex ($B_{i}$) operators. These operations are required to satisfy the fermionic commutation relations, which can be accounted for by utilizing Majoranas $\gamma_i$
\begin{equation}
    \gamma_{2i} = a_{i}^{\dagger}+a_{i}, \quad \gamma_{2i+1}=-i\left(a_{i}^{\dagger}-a_{i}\right)
\end{equation}
as
\begin{equation}
    B_{i} = -i\gamma_{2i}\gamma_{2i+1}, \quad A_{ij} = -i\gamma_{2i}\gamma_{2j}
\end{equation}
The Hamiltonian can then be mapped by replacing each $a_i^{\dagger}a_j$ combination with
\begin{equation}
    a_{i}^{\dagger}a_{j} = \left(1-B_i\right)\tilde{A}_{ij}\left(1-B_j\right).
\end{equation}
where $\tilde{A}_{ij}$ can be any product of $A_{ij}$ such that $\tilde{A}_{ij}=A_{ik}^{a}A^{b}_{kl}..A^{c}_{mn}A^{d}_{nj}$, where the superscript signifies which of the multi-edges is utilized.  The definition of the algebraic properties of $A$ and $B$ is
\begin{equation}
    B_{i}^{\dagger}=B_i, \quad A_{ij}^{\dagger}=A_{ij}
\end{equation}
\begin{equation}
    B_{i}^2 = I, A_{ij}^2=I
\end{equation}
\begin{equation}
    B_i B_j = B_j B_i, \quad A_{ij} = -A_{ji}
\end{equation}
\begin{equation}
    A_{ij}B_{k} = \left(-1\right)^{\delta_{ik}+\delta_{jk}}B_{k}A_{ij}
\end{equation}
\begin{equation}
    A_{ij}A_{kl} = \left(-1\right)^{\delta_{ik}+\delta_{il}+\delta_{jk}+\delta_{jl}}A_{kl}A_ij
\end{equation}
For closed loops on the graph, the $A_{ij}$ satisfy
\begin{equation}\label{eq.stabs}
    i^{p}A_{ij}A_{jk}A_{kl}...A_{mi}=I.
\end{equation}
The loops of Eq. \ref{eq.stabs}  length $p$, which are products of $p$ interactions, are the stabilizers that define the codespace and can be used for error mitigation and correction. Interactions that are not directly connected on the simulation graph can be defined by taking the connected interactions along a path (i.e. $A_{jl}=A_{jk}A_{kl}$). There are certain cases where not taking the direct route results in a lower term-weight or increases parallelization.\cite{Bringewatt2023parallelization} 

In the GSE, $A_{ij}$ and $B_{i}$ are defined using local Majoranas\cite{Setia_2019}. These local Majoranas define the form of the mapped Hamiltonian and the corresponding stabilizers. There are many properties that one could optimize for when choosing these local Majoranas. For example, the local Majoranas defined in Ref \citenum{Setia_2019} are a good choice if one is interested in having a certain code-distance. If one is interested in reducing the average term weight, the JKMN (ternary-tree) mapping\cite{JKMN} is a good choice. If not otherwise specified, we utilize the Majoranas of Ref \citenum{Setia_2019} as these are the most likely to be useful beyond near-term quantum computers as a part of a fault-tolerant framework.

\section{Techniques to improve results}
To obtain the best results with the GSE, we utilize a variety of techniques that do not depend on the structure of the Hamiltonian.  Using these techniques, the circuit depth, number of qubits, properties of the hardware, and error detecting properties can be tailored to obtain good results across a wide variety of chemical systems. Some of these techniques can be applied to any fermion-to-qubit mapping and we will highlight when this is the case.

\subsection{Removing edges}

\begin{figure}
    \centering
    \includegraphics[width=0.5\linewidth]{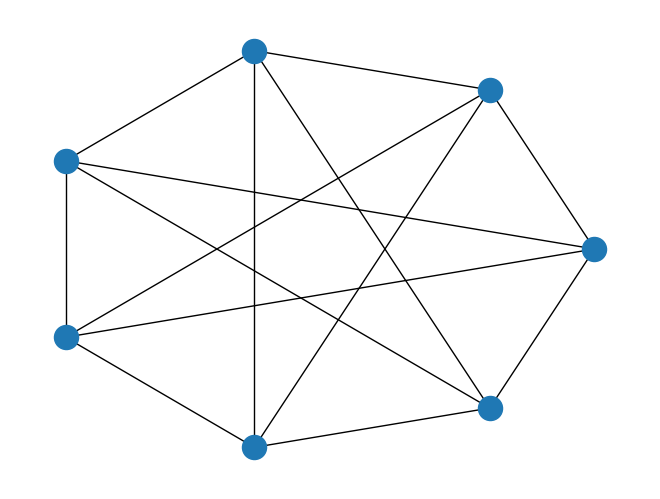}
    \caption{An example of a distance-2 graph for 7 modes where each excitation is a maximum of 2-hops away.}
    \label{fig:7mode_distance2}
\end{figure}

In general, all molecular Hamiltonians are fully connected within a spin-sector so the number of qubits $N$ required is $2 \genfrac{(}{)}{0pt}{2}{M}{2}$ when following the original GSE\cite{Setia_2019} exactly. However, it is always possible to remove edges from simulation graph as outlined in Refs \citenum{Chien2020, Chien2022, Chien2023} to reduce the number of required qubits. For the GSE, the degree $d$ of the graph determines the code-distance (for suitably chosen local Majoranas\cite{Setia_2019}) and the number of qubits is $\frac{d}{2}\times N$. As an example of pruning interaction edges, if one has seven spatial orbitals, a fully connected graph would have weight six and require $21$ qubits per spin sector. This can be  reduced to a weight-4 graph (requiring $14$ qubits per spin sector), while keeping the number of hops required for any operation at a maximum of 2 as shown in Figure \ref{fig:7mode_distance2}.

As a side note, using a fully connected graph and the local Majoranas outline in Ref. \citenum{Setia_2019} for a degree-6 fully connected graph (i.e. a molecule with 7 molecular orbitals) is able to detect all degree-two Pauli errors as expected, but additionally, the only undetectable degree-3 operations are the logical $B_{i}$ operators (e.g $Z_{3i}Z_{3i+1}Z_{3i+2}$. Therefore, even though the code is only distance-3, it can detect all non-logical degree-3 errors and has some properties of a distance-4 code.

If one were to fully prune the graph such that the simulation graph is a line with only nearest-neighbor connections with no connections across the graph and is shown in Figure \ref{fig:single_edge}, the result would be the Jordan-Wigner mapping. This would have high-weight Pauli terms for interactions between spatial orbitals that are far apart with the maximum hops required would being $7$. Therefore, it is possible to recover the Jordan-Wigner mapping from the GSE.

\subsection{Multi-Edges \label{sec:multi_edges}}
\begin{figure}
     \centering
     \begin{subfigure}[b]{0.5\textwidth}
         \centering
         \includegraphics[width=\textwidth]{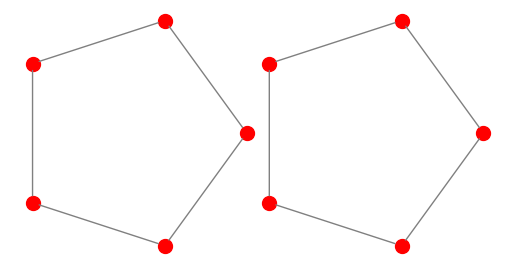}
         \caption{Single Edge}
         \label{fig:single_edge}
     \end{subfigure}
     \hfill
     \begin{subfigure}[b]{0.5\textwidth}
         \centering
         \includegraphics[width=\textwidth]{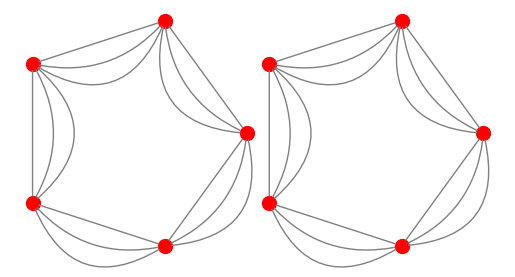}
         \caption{Three edges}
         \label{fig:three_edge}
     \end{subfigure}
        \caption{The simplest interaction graph that retains the spin-parity symmetry with single and multi-edges}
        \label{fig:three graphs}
\end{figure}

When designing the simulation graph, it is also possible to add additional edges $E_e$ as described in Ref \citenum{Setia_2019}. The extra stabilizers that are added are the $E_e-1$ self loops. The simplest graph that includes the Hamiltonian's spin-parity stabilizer is a loop that includes all modes and connects the last mode to the first directly, which is shown in Figure \ref{fig:single_edge}. This is similar to the Jordan-Wigner graph but the connection between the first and last mode is absent there. One can easily increase the distance of the code by adding multi-edges as shown in Figure \ref{fig:three_edge}. 

In general, this would increase the weight of all logical operations and the stabilizers generated through the self loops. However, it is possible to choose the Majorana's such that all stabilizers are weight-6 regardless of how many multi-edges are utilized. This is done by ordering the local Majoranas for $d=2k+1$ as 
\begin{equation}\label{eq.tbqcc}
    m=(\underbrace{Z...Z}_{k}Y\underbrace{I...I}_{k}),\, n = (\underbrace{Z...Z}_{k}X\underbrace{I...I}_{k})
\end{equation}
where $(c)$ indicates the $k$-cyclic permutations of the Pauli string. The $A$ operators are then defined as 
\begin{equation}\label{eq.diff_edge}
    A_{ij}^{c}=\underbrace{I...I}_{i\times d}(\underbrace{Z...Z}_{k}Y\underbrace{I...I}_{k})(\underbrace{Z...Z}_{k}X\underbrace{I...I}_{k})\underbrace{I...I}_{(m-j-1)\times d}
\end{equation}
where $c$ signifies the shift in cyclic permutation. The $B_i=\underbrace{I...I}_{i\times d}\underbrace{Z...Z}_{d}\underbrace{I...I}_{(m-i-1)\times d}$. The stabilizers $S_{i,c}$ formed from the multi-edges are defined as 
\begin{equation}
S_{i,c} = A_{i,i+1}^c A_{i,i+1}^{c+1},\quad c=1,..,d-1,
\end{equation}
which always have weight 6 regardless of the distance $d$. This is not the first time that a local fermion encoding has obtained stabilizers with weight independent of code distance. In Refs. \citenum{landahl2023logicalfermionsfaulttolerantquantum,Algaba2025, Wei2025}, they also obtain an increasing code distance with constant stabilizer weight but their derivation comes from topological defect derivation\cite{Algaba2025} or concatenating with a fermionic color code\cite{Wei2025}. Whether those constructions could be combined with the techniques utilized here is an open question. Like most work with local mappings, the discussion in those papers is restricted to simpler Hamiltonians with nearest-neighbor interactions. 

One fortuitious benefit of the construction outlined here is that the overhead of next-nearest neighbor interactions does not increase with code distance. For example, the logical interaction $A_{0n}=A_{01}^{c}A_{12}^{c},,,A_{n-1,n}^c$ for any choice of cyclic permutation $c\in[0,d-1]$, acts only on a single qubit of the $d$ qubits assigned to mode $1$ through $n-1$ with which of the $d$ qubits depends on the path $c$ taken. Therefore, an operation between $n$ next-nearest neighbor requires $n$ additional qubits to be acted on irregardless of $d$. This means that one can judiciously choose the path to implement logical operations such that operations using this simple GSE mapping can be applied in parallel that are impossible using standard Trotter gadgetry\cite{Whitfield_2011} in (for example) Jordan Wigner. To find this parallelization, one could possibly utilize the gadgetry outlined in Ref. \citenum{Bringewatt2023parallelization}.

\begin{figure}
     \centering
     \begin{subfigure}[b]{0.7\textwidth}
         \centering
         \includegraphics[width=\textwidth]{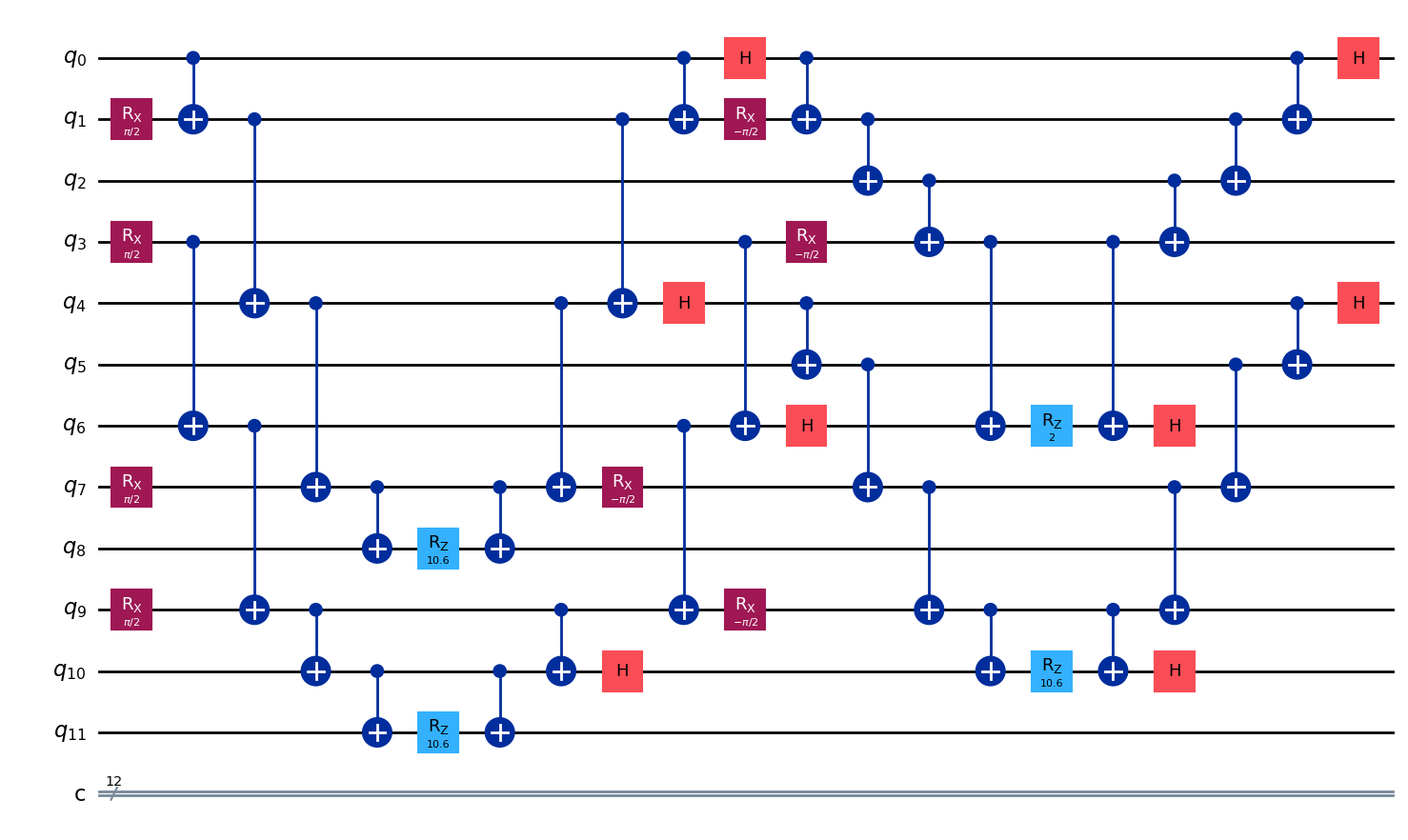}
         \caption{GSE distance 3}
         \label{fig:small_h_d3}
     \end{subfigure}
     \hfill
     \begin{subfigure}[b]{0.7\textwidth}
         \centering
         \includegraphics[width=\textwidth]{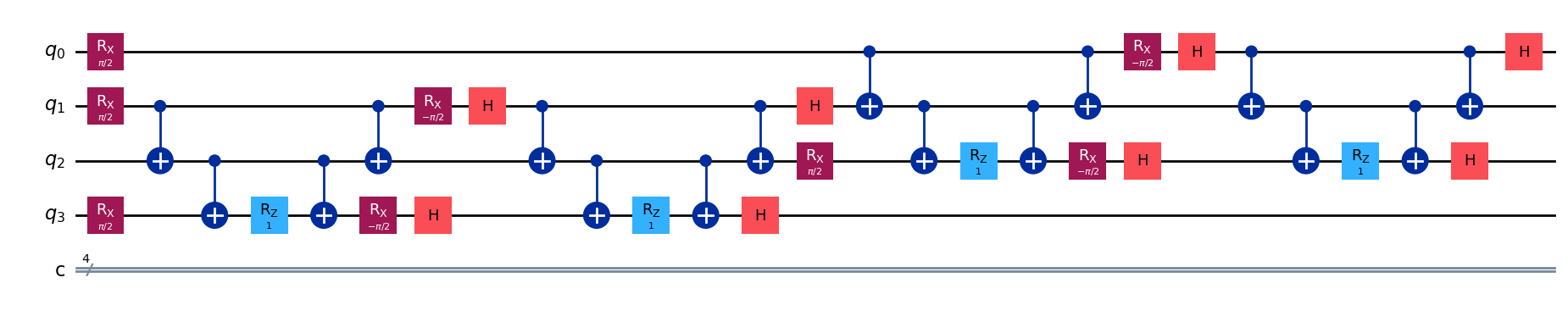}
         \caption{Jordan Wigner}
         \label{fig:small_h_jw}
     \end{subfigure}
        \caption{The Trotter-Suzuki circuit for the Hamiltonian $a_0^{\dagger}a_2+a_1^{\dagger}a_3+h.c.$ mapped using both GSE distance-3 with multi-edges and Jordan-Wigner. The resulting circuit depth for the distance-3 GSE is 21 while the Jordan-Wigner circuit depth is 27. By carefully choosing which path to define an excitation, it is possible to reduce the circuit depth.}
        \label{fig:shorter_depth}
\end{figure}

As a concrete example, the Hamiltonian $a_0^{\dagger}a_2+a_1^{\dagger}a_3+h.c.$ maps to terms $A_{02}B_2+A_{13}B_3+B_0A_{02}+B_1A_{13}$. Using the distance 3 GSE, the term $A_{01}^{0}A_{12}^{0}B_2=-Z_0 Y_1 Z_4 Y_7 Z_8$ has no overlapping support with the operation $A_{12}^2A_{23}^2B_3=-1 Y_3 Z_5 Z_6 Y_9 Z_{10}$. Similarly, $B_0A_{12}^1A_{23}^1$ has no overlapping qubits with $B_1 A_{12}^2A_{23}^2$.  When utilizing the Jordan Wigner mapping, these terms can never be applied simultaneously as $B_i=Z_i$ and $A_{ij}=X_{i}Y_{j}$ so $A_{01}A_{12}B_2=-X_{0}Z_1X_{2}$ and $A_{12}A_{23}B_3=-X_1Z_2X_{3}$.  Using standard ladder operations\cite{Whitfield_2011}, the GSE mapped Hamiltonian has a circuit depth of 21 while the JW mapped operations have a depth of 27. The circuits for these are shown in Figure \ref{fig:shorter_depth}. Therefore, the circuit depth can be reduced while having error correcting/detecting properties. In this way, it is clear that there is not always a trade-off between error detection and circuit depth when using the GSE instead of Jordan-Wigner.

Finally, we also note that the GSE of Eq. \ref{eq.tbqcc} can be recognized as a tail-biting quantum convolutional code\cite{Forney2007}. This means that it is not unreasonable to presume that there is a scalable path to decoding errors in a fault-tolerant manner using the Quantum Viterbi decoder\cite{Chau1999}, but this is beyond the scope of this work.

\subsection{Reducing term weight}
After a mapping has been made for chemical systems, there are, in general, more terms in the mapped qubit Hamiltonian than in other mappings due to terms differing by the product of a stabilizer\cite{Chien_2019, Chien2023}. For example, if one maps $a_1^{\dagger}a_0$ and chooses different paths from Eq \ref{eq.diff_edge}, (i.e. different $c$) for the term $A_{01}^c$, then the resulting qubit operator is different, but the logical operation is the same. This does not present itself for simpler Hamiltonians but we can be utilize this property to reduce the average weight of the terms in the Hamiltonian after mapping. By using the mapped logical operation  with the lowest-weight Pauli terms to represent the desired logical operation, we can obtain a mapped Hamiltonian with competitive Pauli weights to standard mappings but with the error correcting properties still present.

During the initial mapping procedure, the shortest path between modes is always taken to obtain the mapped terms that are more numerous then the original Hamiltonian. To locate the identical logical operations, we transform the mapped Hamiltonian so that the first $S$ qubits represent the transformed $S$ stabilizers and the $2N-S$ qubits represent the logical operations. This transformation can be found by defining the Clifford tableau through the list of stabilizers followed by the vertex operators\cite{gidney2021stim}. The resulting transformed operator has terms with $Z,I$ on the first $S$ qubits while the other qubits are any of $I$, $X$, $Y$ or $Z$. What combination of $Z$ and $I$ act on the stabilizer qubits determines the length of the Pauli string in the corresponding standard form (after transforming back to the original representation). Any term that has the same operations on the last $N-S$ qubits is an equivalent logical operation. By keeping track of which standard term is transformed to a particular logical operation, a reasonable approximation to the shortest Pauli length standard GSE type operation can be found by converting all logical terms to that standard-form shortest weight operator (i.e. by converting the first $S$ qubits to the same product of $Z,I$). By combining logical operations into terms that have the shortest Pauli weight, one can ensure that terms that should cancel do cancel, and the resulting logical operators are reasonably short in polynomial time. 

As an example, two different techniques to map the $A_{02}$ operation using a singly-connected loop (i.e. figure \ref{fig:single_edge}), are $A_{01}A_{12}$ and $A_{04}A_{43}A_{32}$. 
These operations differ only by a product of the stabilizer $A_{01}A_{12}A_{23}A_{34}A_{40}$, which is the logical $I$ and can also be identified as the spin-parity symmetry. Therefore, after transforming to the form described above, depending on the transformation circuit, one of the operators will have $Z$ on the first qubit and the other $I$ with the remaining non-stabilizer qubits being identical.

It should be noted that this utilization of the spin-parity symmetry could immediately be used to reduce the maximum weight of the Jordan-Wigner mapped Hamiltonian to $M+2$ instead of $2M$ as long as a predefined parity of the electron number for each spin is used. If the number of $Z$ operators in the alpha (beta) spin-sector is greater than $M/2$, multiply that term by $\pm Z_{0}...Z_{M-1}$ ($\pm Z_{M}...Z_{2M-1}$) where the sign is determined by the spin-parity of the chemical system examined. After the transformation, the lowest-energy eigenvalue of the transformed Hamiltonian will now (in general) not be the ground state of the chemical system. However, the eigenstate with the correct Hamiltonian stabilizers (i.e. $\pm Z_{0}...Z_{M-1},\,\pm Z_{M}...Z_{2M-1}$) will be the correct energy for the physical system examined but with a max-Pauli weight of $M+2$

\subsection{An alternative to direct stabilizer measurements\label{sec:rotation}}
Throughout the discussion above, we have focused on GSEs that have error detection/correction properties that can be used to mitigate errors. The discussion below explores how to exploit these properties effectively when mid-circuit measurements are not available. The error detection we utilize here does not preclude the use of other error mitigation techniques\cite{Cai_2023} but we focus only on post-selection applied to the logical encoding of fermions in the GSE.

In order to obtain optimal error mitigation results, we do not utilize standard stabilizer measurements circuits\cite{Hagge2023} with mid-circuit measurements, but instead create circuits that transform all stabilizers to the $Z$-basis and measure all operators and stabilizers simultaneously. Although one could easily generate the circuit that transforms the requested fully commuting logical operators $\bar{P_0},\bar{P_l}$ and stabilizers $S_{0}...S_{N-L+2}$ to $Z_{0}...Z_{N}$, this is not optimal as the code distance is reduced to $1$. Instead, we utilize the process 1) Generate a code space state, 2) simulate (using stim\cite{gidney2021stim}) a set of single-ancilla stabilizer measurements for the set $\bar{P_0},...,\bar{P_l},S_{0}...S_{N-L+2}$. This will project the logical and stabilizer state into a new code space which can then be used to generate the rotation circuit. 3) Generate the circuit that would create this new state (regardless of the simulated measurements) from the zero state and apply the inverse of that to the end of the state-preparation circuit. Apply the same circuit to the fully-commuting terms and the stabilizers which now only contain $Z$.  The circuit depth of this transformation will be $N^2$. As an illustrative example, consider the simplest GSE code containing 4-electrons in 3 orbitals. The interaction graph is two triangles with stabilizers $Z_0Z_1Z_2,\, Z_3Z_4Z_5$ (which in this case is the same as the Jordan Wigner mapped stabilizers).

\begin{figure}[ht]
    \centering
    \includegraphics[width=0.4\linewidth]{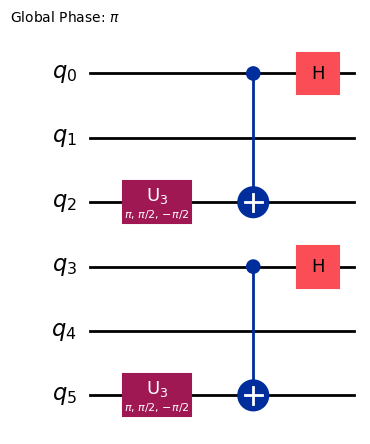}
    \caption{The spin-separated rotation circuit that changes the canonical stabilizers $XIXIII$, $-ZIZIII$, $IZIIII$, $IIIXIX$, $-IIIZIZ$, and $IIIIZI$ to  all $Z$ operators.\label{fig:rotation_circuit}}
    
\end{figure}

The above example utilizes a fully-commuting algorithm across both spin-sectors\cite{Yen2023}. In order to keep the rotation circuit depth smaller, we modify the fully commuting algorithm to ensure that operations commute across spin-sectors as opposed to operations across both. This results in an average circuit depth of this post-selection circuit to be $\mathcal{O}(N^2/4)$ as opposed to $\mathcal{O}(N^2)$ if fully commuting groups were used. If we wish to measure the terms that commute with the canonical (as defined in the Tableau class of stim\cite{gidney2021stim}) stabilizers $XIXIII$,$-ZIZIII$,$IZIIII$,$IIIXIX$,$-IIIZIZ$, and $IIIIZI$ then the circuit that rotates all operators to $Z$ only while keeping the stabilizers all $Z$ ( in this case $Z_1Z_2, \, Z_4Z_5$) is shown in Figure \ref{fig:rotation_circuit}. 

\begin{figure}
    \centering
    \includegraphics[width=0.9\linewidth]{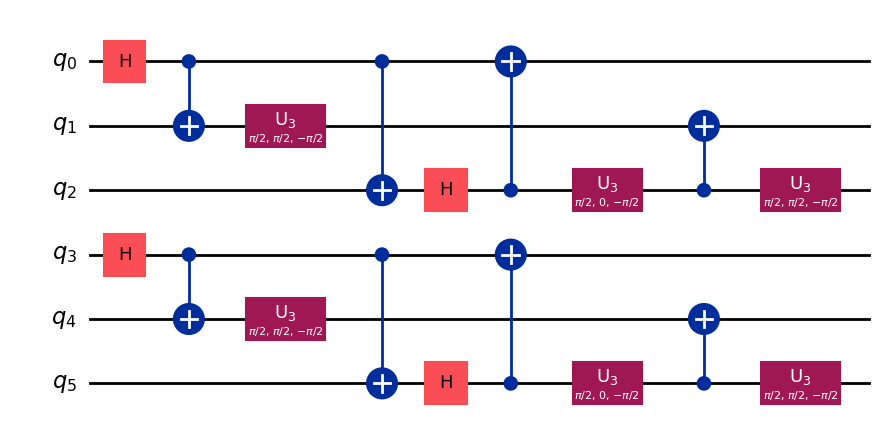}
    \caption{The circuit that rotates the same operators such that the stabilizers are changed to $Z_1,\, Z_4$.}
    \label{fig:standard_rotation_circuit}
\end{figure}

Without using this procedure, one could also directly determine the circuit that takes a set of operators that determines the logical space (in this case $XIX$ and $YIY$) for each spin and determine the circuit that rotates each to $Z_0 Z_1 Z_2$. This would result in the circuit given in Figure \ref{fig:standard_rotation_circuit}. As can be seen, there are now $4$ CNOT gates and a depth that is much longer.
When comparing the operators transformed, the technique used in this manuscript creates a circuit that $10$ of $12$ transformed operators having overlapping support with the transformed stabilizers while for the standard technique only $4$ of the $12$ have overlapping support.


\section{Results}

\subsection{Reduction of Circuit Depth}
In order to analyze the circuit depth resulting from GSE vs standard mappings, we utilize a graph coloring algorithm\cite{Bringewatt2023parallelization} to find groups of Pauli strings in the mapped that act on different qubits. By multiplying the number of groups by the average Pauli string length, we get a reasonable approximation to the average circuit depth required to perform a single first-order trotter step. For the purposes of this analysis, we use the H$_2$O molecule in a cc-pVTZ basis with a restricted active space such that the number of orbitals $M$ grows.
\begin{figure}
     \centering
     \begin{subfigure}[b]{0.45\textwidth}
    \centering
    \includegraphics[width=0.9\linewidth]{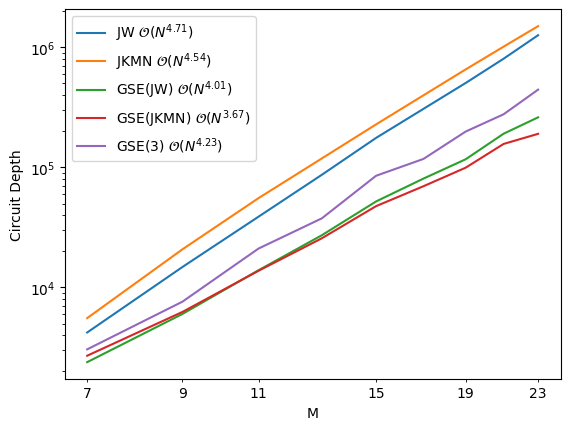}
    \caption{Different Mappings}
    \label{fig:circuit_depth}
     \end{subfigure}
     \hfill
     \begin{subfigure}[b]{0.45\textwidth}
    \centering
    \includegraphics[width=0.9\linewidth]{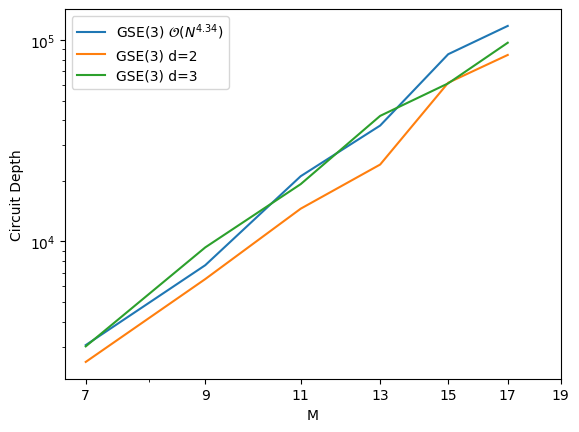}
    \caption{Different $d$}
    \label{fig:enter-label}
     \end{subfigure}

        \label{fig:circuit_depth_comparison}
            \caption{The scaling of the circuit depth for different distances of code. As can be seen, the circuit depth scales similarly regardless of graph simulated.}
\end{figure}

As the number of qubits scales as $M^2-M$ when simulating directly on the fully-connected interaction graph of the Hamiltonian, it is important to note that similar scaling of circuit depth is noted when using a reduced-weight graph similar to Figure \ref{fig:three_edge} where the number of qubits is $Md$.

To exemplify the improvements made with the above techniques, we compare to the results of the original superfast encoding (OSE)\cite{Setia_2018} for molecules. Ref \citenum{Chien_2019} showed that the OSE was worse than JW for all molecules tested. This is no longer the case with the improvements outlined in this manuscript. For the propyne molecule with 19 molecular orbitals, we show in Table \ref{tab:my_label} that GSE is now better than JW and OSE on all metrics aside to the number of qubits. If Ref \citenum{Chien_2019} had access to these tools, their metrics would have been similar but the original superfast encoding could not generate an arbitrary code distance.
\begin{table}[]
    \centering
    \begin{tabular}{c|c|c|c}
         Attribute & JW & OSE\cite{Chien_2019} & GSE(JKMN) \\
         \hline
         Average Term Weight & 12.85 & ~35 & $\boldsymbol{11.88}$ \\
         Max Term Weight & 38 & 72  & $\boldsymbol{16}$ \\
         $\#$ of qubits & $\boldsymbol{38}$ & 342 & 342 \\
         Circuit Depth & 5.1$\times 10^6$ & - & $\boldsymbol{1.2\times 10^6}$
    \end{tabular}
    \caption{The comparison on various attributes for the propyne molecule from the AE6 dataset for OSE, JW and GSE(JKMN). GSE does better on all attributes except for $\#$ of qubits.}
    \label{tab:my_label}
\end{table}

\subsubsection{[[2N, N, 2]] GSE code}
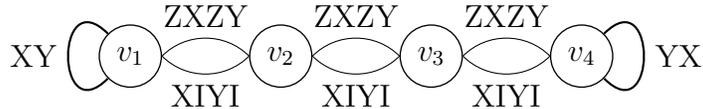
\begin{figure}
    \centering

\begin{tikzpicture}
    \SetVertexMath      
    
    \Vertex[x=0,y=0]{v_1}
    \Vertex[x=2,y=0]{v_2}
    \Vertex[x=4,y=0]{v_3}
    \Vertex[x=6,y=0]{v_4}
    

    \draw[-] (v_1.east) to[out=-45,in=-135] node[midway, below] {XIYI} (v_2.west) ;
    \draw[-] (v_1.east) to[out=45,in=135] node[midway, above] {ZXZY} (v_2.west) ;
    \draw[-] (v_2.east) to[out=-45,in=-135] node[midway, below] {XIYI} (v_3.west) ;
    \draw[-] (v_2.east) to[out=45,in=135] node[midway, above] {ZXZY} (v_3.west) ;
    \draw[-] (v_3.east) to[out=-45,in=-135] node[midway, below] {XIYI} (v_4.west) ;
    \draw[-] (v_3.east) to[out=45,in=135] node[midway, above] {ZXZY} (v_4.west) ;
    
    
    \Loop[dist=1cm, dir=EA, label=YX, style=right](v_4)    
    \Loop[dist=1cm, dir=WE, label=XY, style=left](v_1)
        
 \end{tikzpicture}
     \caption{The GSE for a four mode problem. The local majoranas are $ZY,ZX,XI,YI$ and chosen properly ensure that nearest neighbor operations are the minimum weight-2.}
    \label{fig:distance-2-layout}
\end{figure}

In order to show explicitly how GSE can be utilized to decrease the circuit depth of evolving fermionic systems, we now introduce an error-detecting GSE code that can implement the orbital rotation procedure with similar depth and approximately twice as many operations. The graph on which this encoding is generated uses self-loops at the end-points of a linearly connected graph as shown in Figure \ref{fig:distance-2-layout} instead of using the tail-biting procedure of Figure \ref{fig:three_edge}.

 The multi-edge $A_{ij}$ operators are $ZXZY$ and $IXIY$. We can now see that we can take $\gamma_{2i}\gamma_{2i+2}=A_{i,i+1}=X_{2i+1}Y_{2i+3}$ and $\gamma_{2i+1}\gamma_{2i+3}=A_{i,i+1}B_{i}B_{i+1}=Y_{2i}X_{2i+2}$ by traversing different paths between modes. The stabilizers are $Y_{2i}X_{2i+1}X_{2i+2}Y_{2i+3}, \, i\in [0,1,...,N-2]$ along with $X_{0}Y_{1}$ and $X_{2N-2}Y_{2N-1}$. The single excitation operator $f_{i}^{\dagger}f_{i+1}-f^{\dagger}_{i+1}f_i=\gamma_{2i}\gamma_{2i+2}-\gamma_{2i+1}\gamma_{2i+1}$ can be performed for this distance 2 code as $X_{2i}Y_{2i+2}-Y_{2i+1}X_{2i+3}$ where each qubit operator term acts on disjoint qubits. Therefore, the depth of the circuit needed to perform orbital rotations has not increased even though we have formed a logical encoding with code distance 2. The logical zero state of this code can be readily initialized using a single entangling gate (CZ) for each pair of qubits representing the $B_i$ operators. For example, the encoding circuit for the four mode problem above is depicted in Figure \ref{fig:standard_rotation_circuit}.
 \begin{figure}
    \centering
    \includegraphics[width=0.3\linewidth]{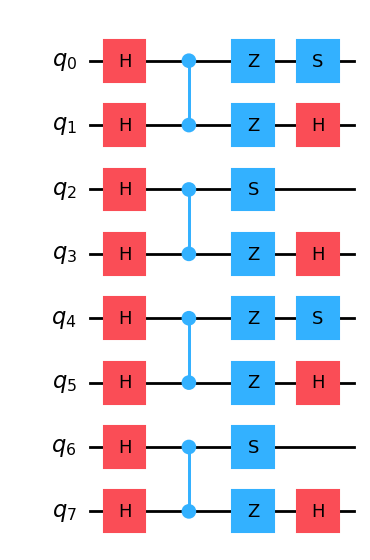}
    \caption{The circuit that encodes the logical zero state for the distance 2 GSE code of Figure\ref{fig:distance-2-layout}}
    \label{fig:vaccuum_circuit}
\end{figure}
 
This results in a logical zero state of $\prod_{j=0}^{N-1}\frac{1}{\sqrt{2}}\left(\ket{10}+i(-1)^j\ket{11}\right)$. By applying the inverse of the vacuum state circuit, one can measure the stabilizers and the orbital occupations simultaneously. The orbital occupations become $Z_{2i+1}, \, i \in[0,N-1]$ while the stabilizers become $Z_{2i}Z_{2i+2}Z_{2i+3},\, i \in[0,N-2]$, $Z_0Z_1$ and $Z_{2N-2}$. This means that we can post-select on the stabilizers being correct to obtain occupation values.

 \subsubsection{Application to dipolar rotors}
\begin{figure}
    \centering
    \includegraphics[width=0.5\linewidth]{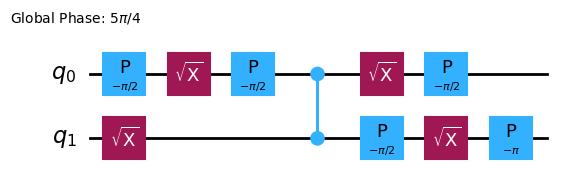}
    \caption{The circuit to transform between even $i,k$ operators of Eq. \ref{eq.rotor_ham} being weight-4 and odd $i,k$ being weight-4}
    \label{fig:transform_gse}
\end{figure}

 In a recent paper\cite{Moeed2025}, it was noted that interacting dipolar rotors could be simulated using the fermionic system
 \begin{gather}\label{eq.rotor_ham}
    H=C- h_0 + h_1 \\
h_0 = \sum_{n=0}^N \sum_{i=nd_m}^{(n+1)d_m}m_i^2 a_i^{\dagger}a_i \\ 
h_1 = \frac{g}{4}\sum_{n=0}^{N-1}\sum_{i=(n+1)d_m}^{(n+2)d_m-1}\sum_{k=(n)d_m}^{(n+1)d_m-1}\big(3 a_i^{\dagger} a_{i+1} a_k^{\dagger} a_{k+1} \nonumber \\ + 3 a_{i+1}^{\dagger} a_{i} a_{k+1}^{\dagger} a_{k} - a_i^{\dagger} a_{i+1} a_{k+1}^{\dagger} a_{k} \nonumber \\ -a_{i+1}^{\dagger} a_{i} a_k^{\dagger} a_{k+1}\big),
\end{gather}
where $N$ is the number of rotors and $d_m$ is the odd number of angular momentum states ([-m,m]) for each rotor. Within each rotor, there are $d_m-1$ electrons (to represent that unary mapping\cite{Moeed2025}), and interactions occur only between nearest-neighbor modes within a rotor. Therefore, the interaction graph is a set of $N$ length-$d_m$ linear graphs. The resulting Hamiltonian only has 2-body terms of $\gamma_{2i}\gamma_{2i+2}\gamma_{2j+1}\gamma_{2i+3}$ and $\gamma_{2i}\gamma_{2i+3}\gamma_{2j+1}\gamma_{2i+2}$. Using the GSE of Figure \ref{fig:distance-2-layout}, the $\gamma_{2i}\gamma_{2i+2}\gamma_{2j+1}\gamma_{2i+3}$ terms have weight-$4$ and multiple terms can be applied in parallel. For $\gamma_{2i}\gamma_{2i+3}\gamma_{2j+1}\gamma_{2i+2}$, one can apply the transformation of Figure \ref{fig:transform_gse} to the circuit for each set of qubits representing a mode and these terms now also have weight $4$ and can be applied in parallel. 

As all two-body terms using the JW mapping also have weight-4, the final result is that the Trotter evolution circuit for the dipolar rotor system can be applied with nearly half the circuit depth and nearly the same number of gates as the JW mapping. The only terms that require more gates is the diagonal $a_i^{\dagger}a_i$ terms which are weight $2$ in GSE and a single pauli-$Z$ rotation in the JW mapping.

\subsection{Error Mitigation}
Putting all the above techniques together, we are now in a position to compare the performance of GSE to the Jordan-Wigner mapping. We use distance 2 (16 qubit) and 3 (24 qubit) codes for simulating noisy qubit coupled cluster type circuit for $(H_2)_2$\cite{DCunha2024}. The code distance was increased by adding multi-edges on the fully connected graph such that each mode has an even number of interactions. As can be seen in Table \ref{tab:h22_gse}, up to distance 3, the measured energy decreases closer to the exact answer. These results were obtained for a noise model that heuristically describes a Quera device which includes circuit level noise and measurement errors. To obtain a noisy correlation energy estimate (i.e. the distance between Hartree-Fock and a correlated wavefunction's energy), we compare the measured energy with zero parameters (which should be the Hartree-Fock energy) and the optimized parameters (i.e. the correlated state). Once again, the accuracy of the answer improves with increasing code distance with the estimated correlation energy much better using GSE compared to the JW encoding. These results demonstrate the effectiveness of GSE to mitigate errors that can improve with increasing code distance.

As another example, we examine the $(H_2)_3$ system which includes 12 spin-orbitals. As can be seen in Table \ref{tab:h22_gse}, the Jordan Wigner mapping has 300mH error in the absolute energy and approximately 200mH error in the correlation energy. The GSE distance-2 code, which required 28 qubits, obtains an absolute error of approximately 180mH and no correlation error with error bars of approximately 16mH. This shows that GSE provides better absolute errors and much better correlation energies for  multiple systems.
\begin{table}[ht!]
    \centering
    \renewcommand{\arraystretch}{1.2}
    \begin{tabular}{cccc}
        \toprule
        Encoding & Optimized Energy& Correlation energy & Measurement $\%$ \\
        \midrule
        \multicolumn{4}{c}{(H$_2$)$_2$}\\
        \midrule
        Jordan Wigner & -1.940$\pm$0.002&-0.0360$\pm$0.002 & $65\%$\\
        GSE Distance 2 & -2.0530$\pm$0.0013& -0.0619$\pm$0.001 & $50\%$\\
        GSE Distance 3 & -2.0772$\pm$0.0269& -0.0699$\pm$0.037 & $20\%$\\
        Exact & -2.1772 & -0.073363 & -\\
        \midrule
        \multicolumn{4}{c}{(H$_2$)$_3$}\\
        \midrule
        Jordan Wigner & -2.566$\pm$0.005&-0.261$\pm$0.008 & $65\%$\\
        GSE Distance 2 & -2.695$\pm$0.011& -0.450$\pm$0.016 & $10\%$\\
        Exact & -2.870 & -0.450 & -\\
        \bottomrule
    \end{tabular}
    \caption{The utility of GSE for the Quera noise model with noise factor of 0.2. The optimized energy is taken using the optimized parameters while the correlation energy is obtained from a noisy estimation of the energy using all zero parameters. As can be seen, the correlation energy estimate is about an order of magnitude greater than the total energy. It can also be seen that GSE is much better at estimating the correlation energy.}
    \label{tab:h22_gse}
\end{table}

\begin{figure}[ht!]
    \centering
    \includegraphics[scale=0.6]{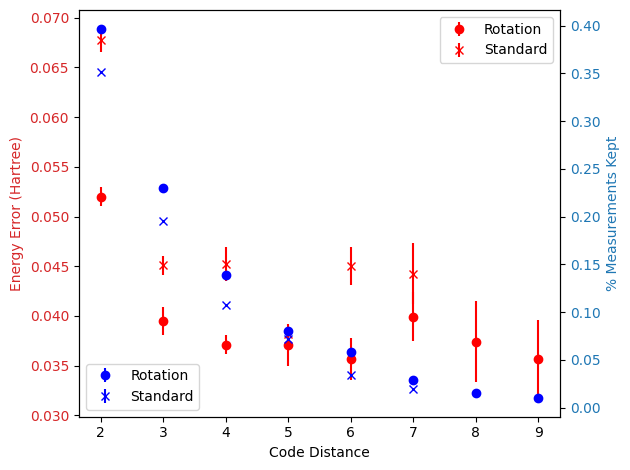}
    \caption{The energy error and measurement cost of the LASSCF circuit for (H$_2$)$_2$ for different code-distances with a fixed shot budget of $10^7$ shots. Each distance uses 8D qubits for a range of simulations from 16 to 72 qubits. The improvement in energy increases until about distance $4$ and then plateaus. The error bars on the energy increases due to fewer measurements being retained for higher distances.}
    \label{fig:las-h2-GSE-stim}
\end{figure}

Table \ref{tab:h22_gse} shows that the zero parameter circuit (Hartree Fock state) is indicative of the performance of the error mitigation properties of GSE. In order to test larger code-distances for the same system, we utilize Stim with a simplified noise model of depolarizing noise (0.001) after every 2-qubit gate. This is a reasonable parameter for state-of-the-art quantum hardware such as that of Quantinuum. The results are shown in Figure \ref{fig:las-h2-GSE-stim} and compare the standard (stabilizer measurement circuits) vs the rotation (described in section \ref{sec:rotation}) method. For both cases, the accuracy in the energy increases until around distance 4 (32 qubits) and then plateaus up to distance 9 (72 qubits). However, the rotation method provides better accuracy compared to the standard stabilizer measurement circuit. Due to fewer measurements satisfying the stabilizers with higher distances, the variance also increases in the measured energy as fewer accepted samples are retained for the given shot budget of 500,000 per group.

\section{Hardware experiment}

Sparse quantum diagonalization has been proposed as a near-term quantum-classical hybrid algorithm that could possibly take advantage of the exponential space available on quantum computers. A key subroutine in many of these papers is the orbital rotation circuit that can be used to perform the single excitations or as a key component of the double excitations (in the LUCJ formalism) from CCSD calculations\cite{Motta2023}. Here we show that the error detecting properties and increased parallelization can be used to obtain superior determinant sampling results on current hardware.

 \subsubsection{Reducing connectivity constraints}
 To perform the orbital rotations in Fermionic space, a device with a square lattice can natively perform orbital rotations using this distance-2 encoding. We will restrict ourselves to real rotations (which are the majority of cases), but the standard orbital rotations can be embedded in the Majorana space as a block diagonal rotation by ordering all even Majoranas followed by odd Majoranas. 
 \begin{equation}
    a_{i}^{\prime} = \sum_{ij}U_{ij} a_j = \sum_{ij}U_{ij} \frac{1}{2}(\gamma_{2j}+i\gamma_{2j+1}^{\prime})=\frac{1}{2}\left( \sum_{ij}U_{ij} \gamma_{2j}+i\sum_{ij}U_{ij} \gamma_{2j+1}\right)
 \end{equation}
 This choice of orbital rotation circuit is optimal if connectivity constraints allow. However, if one is limited to linear connectivity, swap gates are required between each rotation of a given Majorana if it is desired to perform $B_{i}$-type operations at any point (as is done with the LUCJ ansatz). However, as with matchgate circuits\cite{Jozsa2008}, a full set of Majorana rotations is not required to generate a full fermionic Gaussian unitary. In matchgate circuits, the choices are $\gamma_{2i}\gamma_{2i+1}$ and $\gamma_{2i+1}\gamma_{2i+2}$ which in the Jordan-Wigner basis are $Z_{i}$ and $X_{i}X_{i+1}$ respectively. Here, those operations would be $Z_{2i}Z_{2i+1}$ and $Y_{2i+1}Z_{2i+2}Y_{2i+3}$ (or $Y_{2i}Z_{2i+1}Y_{2i+2}$) which are not weight 2 and therefore would require more entangling gates than is optimal. Instead, we order the qubits as $q_{0}q_{1}q_{3}q_{2}...q_{2N-4}q_{2N-3}q_{2N-1}q_{2N-2}$ (that is, reverse the order of every other fermionic mode). We can utilize operations of $\gamma_{4i}\gamma_{4i+2},\, \gamma_{4i+1}\gamma_{4i+3}, \gamma_{2i}\gamma_{2i+1}$ as our basis for performing the general fermionic Gaussian unitary. 

 To elucidate the advantage obtained when using this decomposition, we show the depth of the orbital rotation circuit and the number of entangling operations under linear connectivity. We can see that although the number of entangling gates scales similarly with the number of qubits, the depth now scales with the expected linear fashion.
\begin{figure}
\centering
\begin{subfigure}{0.5\textwidth}
  \centering
  \includegraphics[width=.9\linewidth]{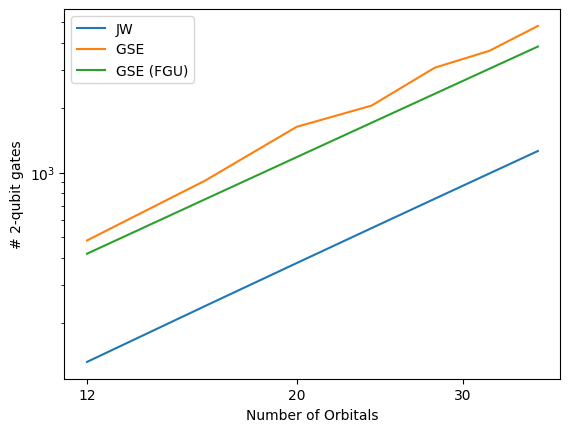}
  \caption{Entangling Gates}
  \label{fig:gates}
\end{subfigure}%
\begin{subfigure}{.5\textwidth}
  \centering
  \includegraphics[width=.9\linewidth]{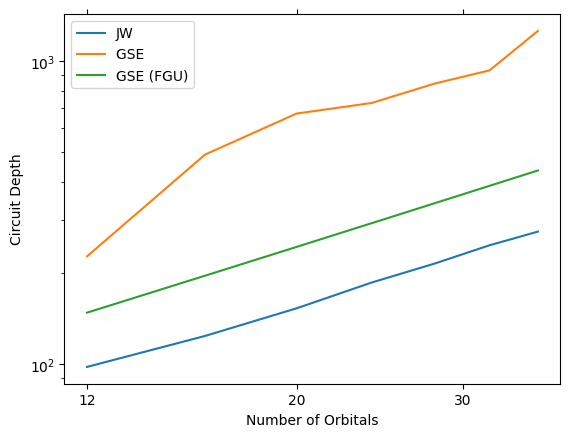}
  \caption{Depth}
  \label{fig:depth}
\end{subfigure}
\caption{The comparison between the GSE (with standard orbital rotation), GSE (FGU) which uses Majoranas and JW. We can see that the depth is greatly reduced when using GSE (FGU) compared to the standard GSE with swap gates required.}
\label{fig:gates_and_depth}
\end{figure}

 Now in the position to test the error-detecting distance-2 GSE code on a platform where connectivity is limited. We apply the techniques to a random unitary for eight orbitals with four electrons on IBM Kingston. We utilize the first N-qubits of the 2N qubits used for the GSE mapping to perform the JW mapping orbital rotation as implemented in ffsim\cite{ffsim} as a comparison. We post-select on both (GSE and JW) for the number of electrons (i.e. 4) but GSE can additionally post-select on the stabilizers. We ran each encoding's orbital rotation circuit for three repetitions (alternating encodings to reduce any effect of hardware drift) such that the number of shots for each such (after post-selctions) was around 10,000 shots. The summary of the results is shown in Table \ref{tab:GSE_vs_JW}. As can be seen, the GSE obtains an occupation difference that has approximately half the RMSE while requiring about twice as many shots.
 \begin{table}[]
     \centering
     \begin{tabular}{c|cccc}
        Encoding  & Qubits & Total Shots & Percentage Accepted & RMSE \\
        \hline
         JW & 8 & $9478 \pm 177$ & $74.8\% \pm 1.5\%$ & $0.0040 \pm 0.0012$ \\
         GSE & 16 & $9877 \pm 1014$ & $38.0\% \pm 3.9\%$ & $0.0020 \pm 0.0002$
     \end{tabular}
     \caption{The RMSE of the occupation numbers for a random orbital rotation of a four electron in eight orbital Fermionic system on IBM Kingston. It can be seen that with approximately twice as many shots, the RMSE error of the recovered occupations is decreased by half using the GSE.}
     \label{tab:GSE_vs_JW}
 \end{table}

 A benefit of using this GSE construction on the heavy-hex architecture of IBM systems is that the fermionic modes are spread out such that the connectivity is increased between spin-sectors. Using the JW mapping, only one in four modes can interact across spins without swaps while this distance-2 GSE can readily interact every other mode.

\section{Conclusion}
In this manuscript, we show that unlike previous work with superfast encodings\cite{Setia_2018,Chien_2019}, it is advantageous to apply GSE to molecular system and not only to highly-structured Fermionic systems such as the Hubbard model. This is done by utilizing a variety of improvements which can be applied to general Fermion-to-qubit mappings but result in GSE performing best for error mitigation.

The first of these improvements is taking advantage of the different paths that the interaction graph of the Hamilton can lead to different weights of the logical operations that result. By judiciously tracking the transformation from codespace-to-logical space of each Hamiltonian, we can choose the lowest-weight Pauli-term that represents that logical operation. Second, we show how multi-edges can be added to the interaction graph that adds error detection/correction properties but does not increase the circuit depth due to the ability of choosing different paths for the interactions to occur. The third improvement is to abandon standard stabilizer measurements in favor of transformations that directly map the desired logical terms and all stabilizers to the $Z$-basis. In order to obtain a good overlap of the resulting stabilizers and the desired logical terms, we simulate a stabilizer measurement process using a Clifford simulator and then obtain the circuit that generates this stabilizer state. This results in better error-mitigation properties while not increasing the circuit depth.

The resulting improvements applied to direct energy estimation of (H$_2$)$_2$ and (H$_2$)$_3$ LASVQE circuits show that much more accurate absolute and correlation energies are obtained compared to the standard Jordan-Wigner encoding. This is especially true for the correlation energy which obtains results that have the correct energy within error-bars of the measured result under currently available hardware noise. We also show that increasing the code-distance up to distance-4 improves the measured energy for the (H$_2$)$_2$ system. 

Although GSE is ideal for hardware that has all-to-all connectivity, we introduce a [[2N,N,2]] GSE that can be easily applied to square-lattice hardware. By utilizing Fermionic Gaussian unitary circuits, instead of standard orbital rotation circuits, we can even use hardware that is (quasi-)linear to generate a logical encoding and perform post-selection. We show that for a random orbital rotation, the RMSE of the measured occupation numbers decreases by approximately half, while requiring about twice as many shots on IBM Kingston. 

In the future, we will look to integrate GSE in a fault-tolerant framework. One of the most promising techniques that we could use is the framework of Ref \citenum{chen2025faulttolerantquantumsimulation} that combines the Trotter evolution of an error-correcting code with another code to implement the CNOT gates of the standard Trotter ladders.  Another technique would be to utilize a hardware-aware circuit compilation\cite{kuehnke2025hardwaretailoredlogicalcliffordcircuits} with flag qubits\cite{Chen2024} to detect errors.

\section{Acknowledgements}
We acknowledge the support of the Welcome Leap foundation under the q4Bio project.

\bibliography{references}
\end{document}